\documentclass[%
 preprint, 3p, twocolumn,
]{elsarticle}

\usepackage{graphicx}% Include figure files
\usepackage{dcolumn}% Align table columns on decimal point
\usepackage{bm}% bold math
\usepackage{float} 
\usepackage{ulem} 
\usepackage{xcolor}
\usepackage{stfloats}
\usepackage{amsmath, amsfonts}
\usepackage{hyperref}% add hypertext capabilities
\usepackage{lineno}
\makeatletter
\def\ps@pprintTitle{%
   \let\@oddhead\@empty
   \let\@evenhead\@empty
   \let\@oddfoot\@empty
   \let\@evenfoot\@oddfoot
}
\def\@@author[#1]#2{\g@addto@macro\elsauthors{%
    \def\baselinestretch{1}%
    \authorsep#2\unskip\textsuperscript{%#1%
      \@for\@@affmark:=#1\do{%
       \edef\affnum{\@ifundefined{X@\@@affmark}{1}{\elsRef{\@@affmark}}}%
     \unskip\sep\affnum\let\sep=,}%
      \ifx\@fnmark\@empty\else\unskip\sep\@fnmark\let\sep=,\fi
      \ifx\@corref\@empty\else\unskip\sep\@corref\let\sep=,\fi
      }%
    \def\authorsep{\space and\space}%
    \global\let\sep\@empty\global\let\@corref\@empty
    \global\let\@fnmark\@empty}%
    \@eadauthor={#2}%
    \g@addto@macro\useauthors{#2; }%
}
\makeatother

\begin{document}

\title{Dark Matter Hot Spots and Neutrino Telescopes}
\author[1,3]{Meighen-Berger S.\fnref{fn1}}
\author[1,2]{Karl M.}

\fntext[fn1]{stephan.meighen-berger@tum.de}

\address[1]{Technische Universit\"at M\"unchen, James-Franck-Stra{\ss}e, 85748, Garching, Germany}
\address[2]{Max Planck Institute for Physics, F\"ohringer Ring 6, 80805 M\"unchen, Germany}
\address[3]{School of Physics, The University of Melbourne, Victoria 3010, Australia}

\date{\today}% It is always \today, today,

\begin{abstract}
We perform a new dark matter hot spot analysis using ten years of public IceCube data. This analysis assumes dark matter self-annihilates to neutrino pairs and treats the production sites as discrete point sources. As a result, these sites will appear as hot spots in the sky for neutrino telescopes, possibly outshining other standard model neutrino sources. Compared to galactic center analyses, we show that this approach is a powerful tool capable of setting the highest neutrino detector limits for dark matter masses between 10 TeV and 100 PeV. This is due to the inclusion of spatial information in addition to the typically used energy deposition in the analysis.
\end{abstract}

%\keywords{Supersymmetry, Cherenkov Telescopes, IceCube, Stau, Neutrino, Navier-Stokes}
\maketitle

\section{Introduction}
With new neutrino telescopes under construction, such as P-ONE \cite{Agostini:2020aar}, KM3NeT \cite{Adrian-Martinez:2016fdl}, GVD \cite{Avrorin:2017tse} and IceCube-Gen2 \cite{vanSanten:2017chb} we expect an improving global sky coverage in the coming years. Especially with collaborative efforts, such as PLE$\nu$M \cite{Schumacher:2021hhm}, the sensitivity to point-like objects emitting neutrinos will increase across the entire sky. With these developments in mind, we perform a hot spot search for self-annihilating dark matter (DM) to study the capabilities of such analyses in the search for new physics.
\\
Multiple dark matter studies using IceCube and or ANTARES have been performed, such as the search for dark matter from the center of our galaxy \cite{IceCube:2015rnn, IceCube:2017rdn, ANTARES:2020leh}. This can be further expanded by including neutrinos produced by diffuse dark matter in our galactic halo or beyond \cite{Arguelles:2019ouk, ElAisati:2017ppn}. Other sources of dark matter signals have been hypothesized and studied as well, such as the sun \cite{Arguelles:2019jfx, Chen:2014oaa, ColomiBernadich:2019upo, Albuquerque:2010bt, Albuquerque:2013xna} and the center of the earth \cite{IceCube:2020wxa}.
\\
In the scenario that dark matter self-annihilates to a neutrino pair, these sources would possess a distinct energy spectrum, a peak at the dark matter's rest mass. They would also appear as "hot spots" in the sky. These would be well-defined regions, as large as the object of interest, where the energy spectrum of the astrophysical neutrino flux would change compared to the diffuse flux \cite{Aartsen:2020aqd}. Especially massive dark matter, with masses above 10 TeV, would cause such a shift. These hot spots would shrink to a point for smaller objects such as suns, planets, or even distant galaxies. See \cite{Murase:2012rd} for an example study on galaxy clusters.
\\
In this letter, we utilize this fact and perform a hot spot  analysis for self-annihilating dark matter directly producing neutrinos\footnote{Code: https://github.com/MeighenBergerS/dmpoint}, using ten years of IceCube public point source data \cite{IceCube:2019cia, IceCube:2021xar}.

\section{Modelling}\label{sec:modelling}
\begin{figure}[t]
    \begin{center}
    \includegraphics[scale=1]{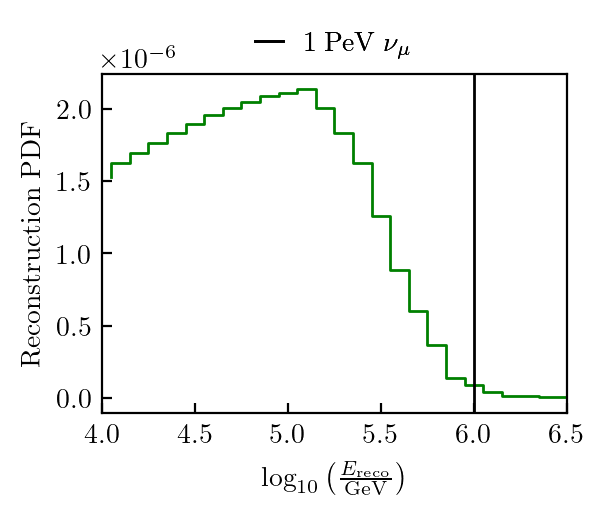}
    \caption{An example showing how a 1 PeV $\nu_\mu$ (black line) is reconstructed according to the published effective areas and mixing matrices. In green, we show the reconstruction pdf which is sampled when generating signal events.}\label{fig:reconstruction}
    \end{center}
\end{figure}

\begin{figure}[t]
    \begin{center}
    \includegraphics[scale=1]{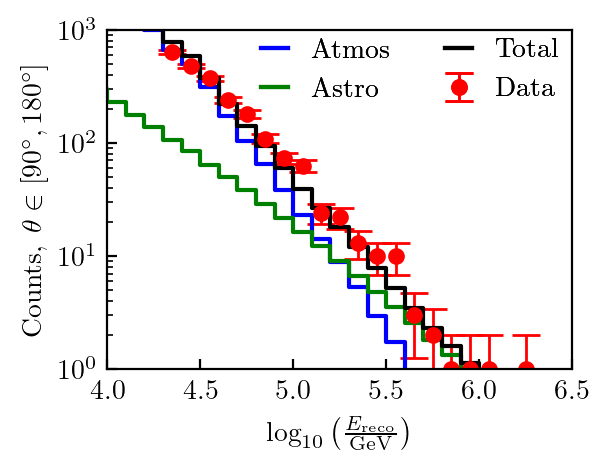}
    \caption{The expected atmospheric and astrophysical counts. We show the individual components in blue (atmospheric) and green (astrophysical). The resulting total counts are given in black, and the measured event counts including statistical errors are shown in red. Here we integrated over $\theta_\mathrm{zenith}$ between $90^\circ$ and $180^\circ$.}\label{fig:atmos_vs_astro}
    \end{center}
\end{figure}

For this search there are two primary backgrounds, the atmospheric neutrino flux, and the astrophysical diffuse flux. To model the atmospheric flux, we use MCEq \cite{Fedynitch:2015zma}, a cascade equation approach. As primary, interaction, and atmospheric models we use H4a \cite{Gaisser:2012zz}, EPOS-LHC \cite{Pierog:2013ria} and NRLMSISE-00  \cite{https://doi.org/10.1029/90JA02125, https://doi.org/10.1029/2002JA009430} respectively. For a comment on other models see \ref{app:shower}. To model the astrophysical diffuse neutrino flux, we use a single power-law
\begin{equation}
    \frac{\mathrm{d}\phi}{\mathrm{d}E} = \phi_0 \times \left(\frac{E}{100\;\mathrm{TeV}}\right)^{-\gamma},
\end{equation}
with $\phi_0 = 1.66^{+0.25}_{-0.27}\;\left[10^{-18}\times\mathrm{(GeV\cdot cm^2\cdot s\cdot sr)^{-1}}\right]$ and $\gamma = 2.53\pm 0.07$ \cite{Aartsen:2020aqd}. We then apply the effective areas and mixing matrices published together with ten years of IceCube data \cite{IceCube:2019cia, IceCube:2021xar}. From this we obtain predicted event counts due to the atmospheric and astrophysical flux. In figure \ref{fig:reconstruction} we show an example of how a 1 PeV neutrino is reconstructed according to the mixing matrix and effective area. The detector will measure a reconstructed muon and its energy $E_\mathrm{reco}$, with the most likely reconstruction energy being approximately at 300 TeV. For this reason even when injecting neutrino lines (in energy) a broad spectrum will be measured. \\
After performing a combined fit on data, where we let the normalization of the fluxes float, we obtain our simulation model for this analysis. The normalization fit results are 1.05 and 1.1 for the atmospheric and astrophysical counts, respectively. The fit is performed on events above 10 TeV and for zenith angles $\theta_\mathrm{zenith} > 90^\circ$. These values correspond to the energy and angular cuts we introduce when analyzing the data.
\\
In figure \ref{fig:atmos_vs_astro} we show the resulting predictions compared to data. We use these simulation values to construct energy weights $\omega_\mathrm{Energy}$ later in the analysis. The weights are the ratio between the expected astrophysical and atmospheric events. The point at which we expect the astrophysical flux to dominate is approximately at 200 TeV, at which point $\omega_\mathrm{Energy} > 1$.
\newline
To model neutrinos produced directly by dark matter self-annihilation in our galaxy we use \cite{Arguelles:2019ouk, Yuksel:2007ac}
\begin{equation}\label{eq:injection}
    \frac{\mathrm{d}\phi_\nu}{\mathrm{d}E_\nu} = \frac{1}{4\pi}\frac{\langle \sigma v \rangle}{2 m_\chi^2}\frac{\mathrm{d}N_\nu}{\mathrm{d}E_\nu}J(\psi),
\end{equation}
with the produced neutrino spectrum $\mathrm{d}N_\nu / \mathrm{d}E$ defined as\begin{equation}
    \frac{\mathrm{d}N_\nu}{\mathrm{d}E_\nu} = 2\delta\left(1 - \frac{E_\nu}{m_\chi}\right)\frac{m_\chi}{E^2}.
\end{equation}
$\langle\sigma v\rangle$ is the thermally averaged self-annihilation cross-section, $m_\chi$ the dark matter mass, $E_\nu$ the resulting neutrino energy, and $J$ is the integral over the target's solid angle and the dark matter density along the line of sight. Here $J$ has units $\mathrm{GeV}^2\mathrm{cm}^{-5}\mathrm{sr}$. We assume Majorana dark matter in the above equation, which can be changed to Dirac DM by dividing the differential flux by 2. We use the above equations to extract limits on the mass $m_\chi$ and the cross-section $\langle\sigma v\rangle$ from the calculated flux limits. See \cite{Aisati_2017} for some possible dark matter models producing lines.

\section{Analysis}
    We bin the events in energy and spatial position to analyze the data. We use a logarithmic grid with ten bins per magnitude for the energy grid. In the case of spatial binning, we construct a linear grid with a step size of $0.2^\circ$, the minimal resolution of the public data set \cite{IceCube:2019cia, IceCube:2021xar}. As mentioned previously, the energy weights $\omega_\mathrm{Energy}$ are the ratio between the expected event counts for the astrophysical and atmospheric neutrinos in each bin. 
    \begin{figure*}[t]
        \begin{center}
        \includegraphics[width=\textwidth]{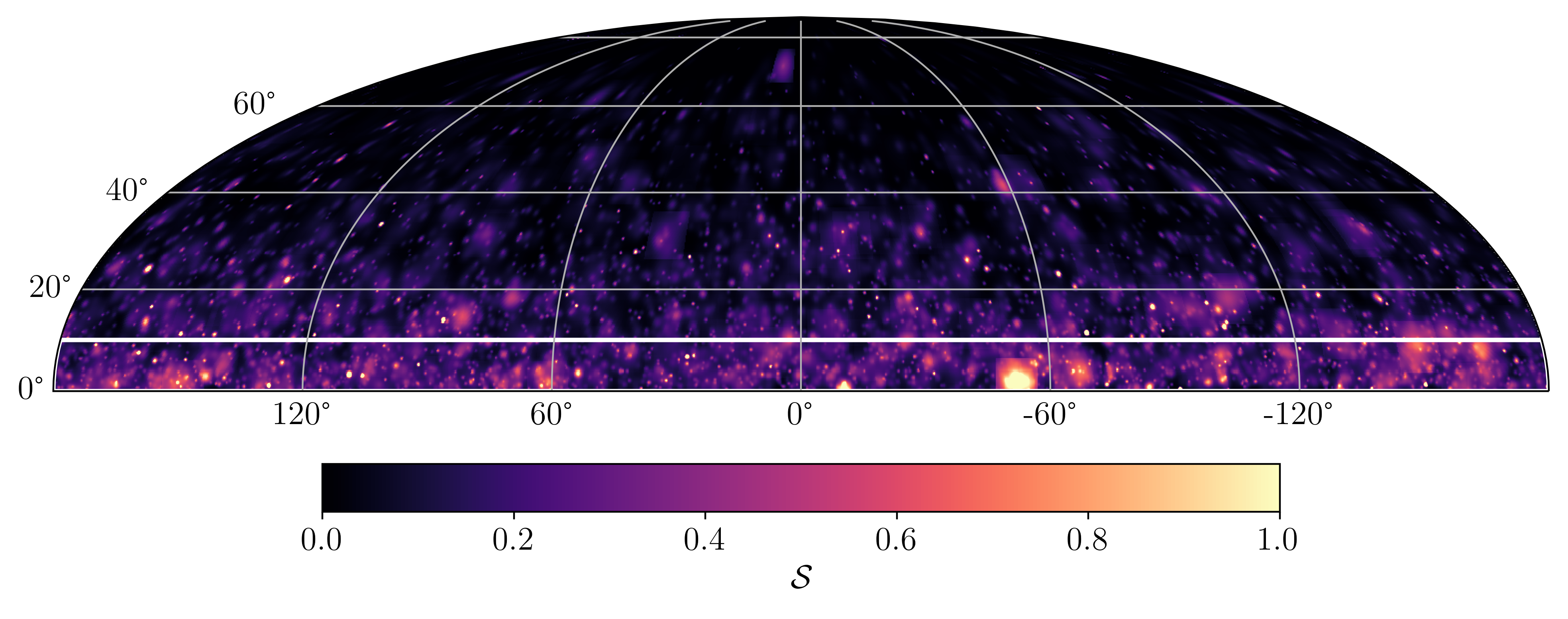}
        \caption{The density sky map in right ascension and declination after weighting and folding with the spatial pdf. The region below the white line is used for this analysis. The plot is made using a Mollweide projection. The color bar describes the score value for each bin. Here we use a cut-off for the maximal score value $\mathcal{S}$ of one.}\label{fig:HeatMap}
        \end{center}
    \end{figure*}
    For the spatial probability distribution function (pdf), we assume a Gaussian distribution $\mathcal{N}$ of the form
    \begin{equation}
    \begin{split}
        \omega_\mathrm{Spatial}^\mathrm{Event}&(\delta,\alpha)  = \\
        & \mathcal{N}(r(\delta^\mathrm{Event}-\delta,\alpha^\mathrm{Event}-\alpha), Error)
    \end{split}
    \end{equation}
    Here $r(\delta, \alpha)$ is the spherical distance between the event's direction and a point on the grid, with $\delta$ and $\alpha$ being the declination and right ascension respectively. $Error$ is the angular reconstruction error of the event. We then assign each spatial grid point, $j$, a score value $\mathcal{S}_j$ according to
    \begin{equation}
        \mathcal{S}_j(\delta,\alpha) = \sum\limits_i\omega_\mathrm{Energy}^i(E_i) \omega_\mathrm{Spatial}^i(\delta,\alpha),
    \end{equation}
    where the sum runs over all events, $i$, in the sample. Figure \ref{fig:HeatMap} shows a heat map of the weighted data events. In white, we show the 10$^\circ$ declination band used in the analysis presented here. For higher declinations, we would expect a reduction in sensitivity. This is due to the attenuation of highly energetic neutrinos in the earth.
    \newline
    For the null hypothesis, we scramble the data in right ascension. For this analysis, we performed $10^5$ scrambles and constructed the cumulative distribution of the score values. From these we derive the mean expectation and standard deviations. See \ref{app:distro} for a more detailed discussion on the distribution.
    \\
    Performing a standard frequentist hypothesis test \cite{Feldman:1997qc, Read:2002hq, James:2000et} on the data distribution, we find no significant deviation between the data and the null hypothesis with a p-value of 0.4. For details on the confidence limit calculation see \ref{app:limits}.

\section{Limits}
\begin{figure}[htb]
        \begin{center}
        \includegraphics[scale=1]{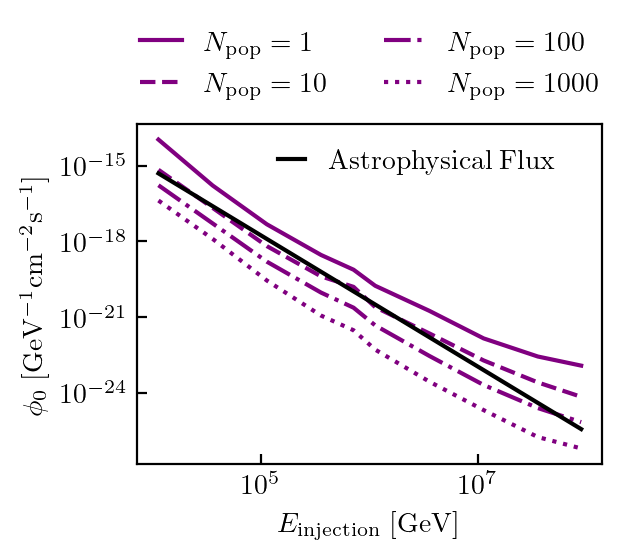}
        \caption{90\% CL limits for the differing population numbers. Here we show the absolute flux limit for a given injection energy. The solid black line depicts the astrophysical diffuse flux as a comparison. It is given in $\mathrm{GeV^{-1}cm^{-2}s^{-1}sr^{-1}}$ and true neutrino energy. The extracted flux limits are depicted using the purple solid ($N_\mathrm{pop}=1$), dashed ($N_\mathrm{pop}=10$), dotted-dashed ($N_\mathrm{pop}=100$) and dotted ($N_\mathrm{pop}=1000$) lines.}\label{fig:flux_limits}
        \end{center}
    \end{figure}

\begin{figure}[htb]
        \begin{center}
        \includegraphics[scale=0.5]{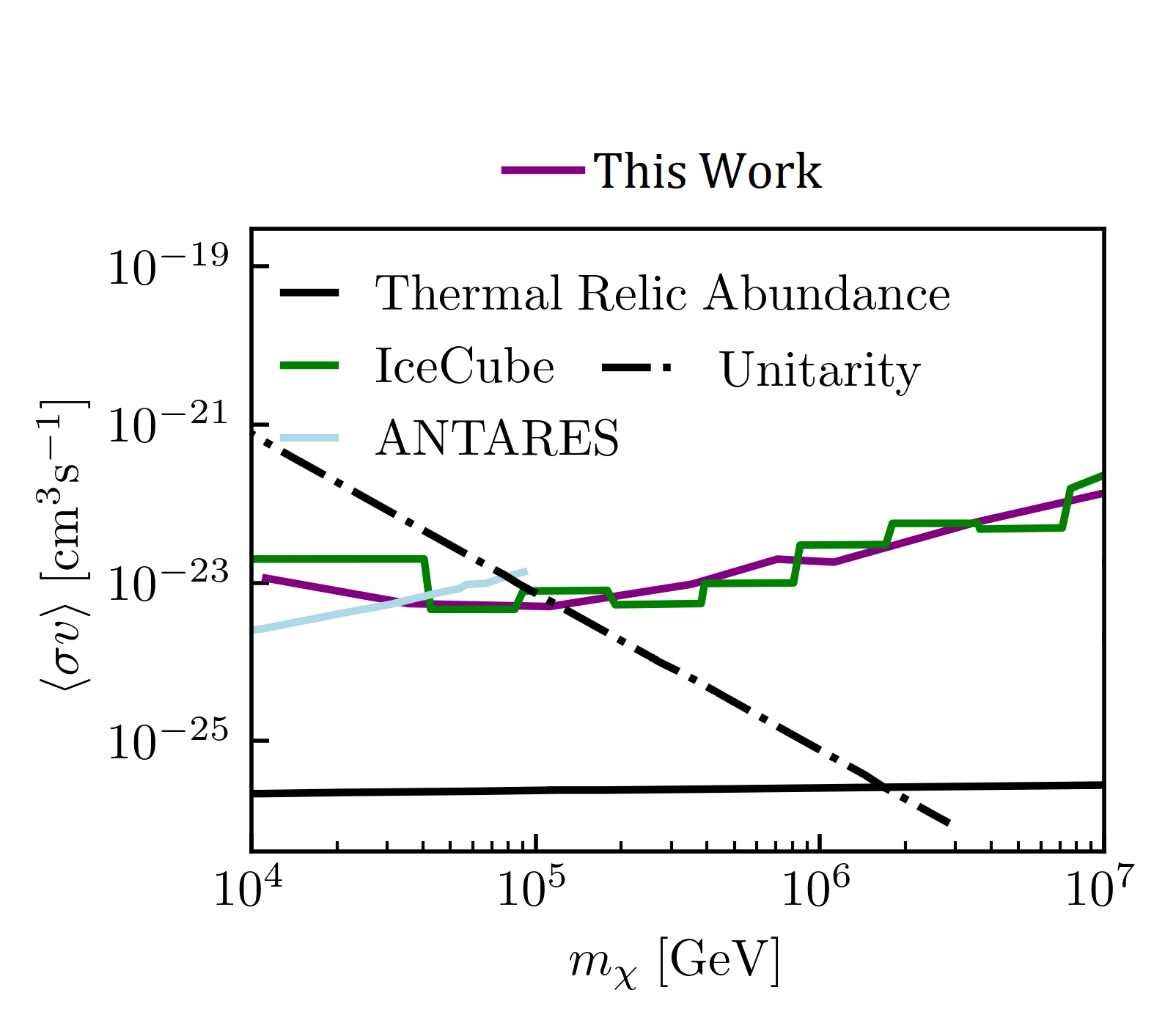}
        \caption{DM limits assuming $J(\psi) = 10^{24}\;\mathrm{GeV^{2}cm^{-5}sr}$ for point-sources in the southern sky. Here we compare the limits to those estimated for extra-galactic emissions in \cite{Arguelles:2019ouk} for IceCube (green) and those measured by ANTARES on the galactic center \cite{ANTARES:2015vis, Albert:2016emp} (light blue). We show the extracted limits using the purple solid ($\mathrm{This\; Work}$) line. We include the constraints set by unitarity (black, dashed-dotted) on non-composite dark matter \cite{Smirnov:2019ngs, Arguelles:2019ouk} and the thermal relic abundance (black, solid) \cite{Arguelles:2019ouk}. }\label{fig:flux_limits_dm}
        \end{center}
    \end{figure}

    To keep the constraints as model independent as possible, we first model signal energy fluxes $\phi_\mathrm{Sig}$ of the form
    \begin{equation}
        \phi_\mathrm{Sig} = \phi_0 \delta(E-E_\mathrm{injection}).
    \end{equation}
    We scan the signal energy, $E_\mathrm{injection}$, and set constraints on the flux normalization $\phi_0$. We assume the angular uncertainty on the signal events lies between 0.5$^\circ$ and 1$^\circ$ degrees. Together with equation \ref{eq:injection} and the modelling procedure described in section \ref{sec:modelling}, we construct the expected events. Here we analyze four different population hypotheses: 1, 10, 100, and 1000 sources distributed randomly in the area of interest. For each hypothesis, we run $10^5$ simulations, letting the flux normalization for the atmospheric and astrophysical background float in the ranges [1, 1.1] and [1., 1.2] respectively\footnote{These ranges are based on the best fit results from Section \ref{sec:modelling}}. Then we construct the 90\% confidence limit. Figure \ref{fig:flux_limits} shows the resulting limits compared to the astrophysical diffuse flux. For a single source, it may emit more than we expect from the diffuse flux, while for $N_\mathrm{pop} > 10$, each object needs to emit less. The ratio between these limits is nearly linear.
    \\
    Until now, we have assumed that we are searching for sources emitting spectral neutrino lines. A natural choice to compare our current results with are those derived for the galactic center. To make this comparison, we need to construct limits on dark matter specific to equation \ref{eq:injection} and set $J(\psi)$. From \cite{IceCube:2015rnn, Yuksel:2007ac} we can use an estimate for an NFW profile \cite{Navarro:1995iw} of $10^{24}$ $\mathrm{GeV^2 cm^{-5} sr}$, using an angular reconstruction error of 1$^\circ$. Utilizing the southern sky sample of the dataset, we set a line source at the position of the galactic center ($\delta\approx -30^\circ)$. We then construct flux limits following the previous discussions and convert to ones on the thermally averaged self-annihilation cross-section $\langle\sigma v\rangle$ and the dark matter mass $m_\chi$. Figure \ref{fig:flux_limits_dm} shows the results. We included the limits set by ANTARES \cite{ANTARES:2015vis, Albert:2016emp} for the galactic center, and the sensitivities estimated for extra-galactic emission in \cite{Arguelles:2019ouk} for IceCube as a comparison. The limit we set here for a single point source is comparable to previous dedicated ANTARES studies and the diffuse extra-galactic emission sensitivity estimate. Note that we present here the estimated sensitivity for the southern sky, the region where the galactic center would appear for IceCube. The sensitivity would be approximately a factor 20 better when searching for sources in the northern sky \cite{IceCube:2019cia}. The limit scales linearly with the $J$-factor.

\section{Conclusion}

We presented a new search for dark matter, including spatial information in addition to the typically analyzed energy distributions. By employing a hot spot population analysis on ten years of public IceCube data, we set limits on self-annihilating dark matter over a broad energy range. We compared the method to previous limits on dark matter in our galaxy as well as estimations on future diffuse extra-galactic sensitivities. In our study, we set comparable limits to both neutrino telescope analyses on self-annihilating dark matter with masses between 10 TeV and 10 PeV. Unlike the previous analyses, the method presented here is agnostic concerning the position and type of object and can be easily extended to regimes beyond self-annihilation. In the future, we expect this method only to improve. Besides the new neutrino telescopes currently being constructed, such as KM3NeT, GVD and P-ONE, this method would only benefit once an actual neutrino source has been discovered. With the evidence of a point source already found with TXS 0506+056 \cite{IceCube:2018cha}, we expect this to happen in the coming years. For this reason, we suggest using hot spot searches to probe for new physics and dark matter.

\section{Acknowledgements}
 The material presented in this publication is based upon work supported by the Sonderforschungsbereich Neutrinos and Dark Matter in Astro- and Particle Physics (SFB1258).

 \nocite{*}
\bibliographystyle{apsrev}
 \bibliography{dm_hotspot}% Produces the bibliography via BibTeX.

\appendix

\section{Atmospheric Shower Differences}\label{app:shower}
To understand the dependence of this analysis on the primary and interaction models used in simulating atmospheric showers, we perform the analysis with other models as well. In the case if the interaction model, we also test Sibyll 2.3c \cite{Riehn:2017mfm}, QGSJET-II \cite{Ostapchenko:2010vb} and DPMJET-III \cite{Roesler:2000he}). For the primary models we use H3a \cite{Gaisser:2012zz} as well as and Gaisser-Stanev-Tilav Gen 3 and 4 \cite{Gaisser:2013bla}. Running the analysis, we found the differences for the final results negligible. The reason for this is the fit to data we perform in the analysis described in section \ref{sec:modelling}. 

\section{Cumulative Distribution}\label{app:distro}
The cumulative distribution function of the score value $\mathcal{S}$ used in the evaluation described above. The background model (black) was constructed by scrambling the data $10^5$ times in right ascencion. The green and yellow regions mark the background model's one and two $\sigma$ confidence intervals.
\begin{figure}[htb]
    \begin{center}
    \includegraphics[scale=1]{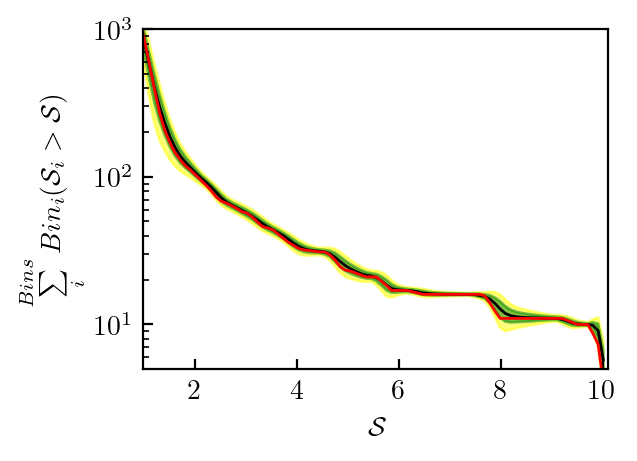}
    \caption{The measured cumulative distribution (red) compared to the mean from the scrambled ones (black). In green and yellow, we respectively show the $1\sigma$ and $2\sigma$ confidence intervals of the scrambled sets. Note that the data agrees well with the null hypothesis. This means there are no significant deviations from a diffuse background assumption.}\label{fig:cumulative}
    \end{center}
\end{figure}
The distribution $\mathcal{D(\mathcal{S})}$ itself is constructed by counting the number of spatial grid points (bins) with a score value $\mathcal{S}_i$ above a given value $\mathcal{S}$
\begin{equation}
    \mathcal{D(\mathcal{S})} = \sum\limits_i^{Bins}Bin_i(\mathcal{S}_i > \mathcal{S}).
\end{equation}

\section{Limit Setting}\label{app:limits}
    To construct the limits on the new flux we perform a frequentist analysis using the CL(s) technique \cite{Read:2002hq, James:2000et}. $CL_s$ is defined as the ratio of the signal + background hypothesis $CL_{s+b}$ and the background only hypothesis $CL_b$ (the left sided p-values of the test statistic distribution):
    \begin{equation}
        CL_s = \frac{CL_{s+b}}{CL_{b}}.
    \end{equation}
    From this the confidence limit is defined as
    \begin{equation}
        1 - CL_s \leq CL.
    \end{equation}
    We define the test statistic $Q$ as
    \begin{equation}
        Q = \frac{\prod\limits_{i=1}^{N_\mathrm{bins}}\mathcal{P}(n_i, s_i+b_i)}{\prod\limits_{i=1}^{N_\mathrm{bins}}\mathcal{P}(n_i, b_i)},
    \end{equation}
    where the product runs over the bins and $\mathcal{P}$ is the Poisson probability mass function. $n_i$, $s_i$ and $b_i$ are the observed signal and background score values respectively. To calculate $CL_s$ this test statistic needs to be constructed for the signal + background and background-only hypothesis. The background expectations $b_i$ are constructed using the mean of the $10^5$ data scrambles.

\section{Galactic Center}\label{app:gc}
\begin{figure}[H]
        \begin{center}
        \includegraphics[scale=1]{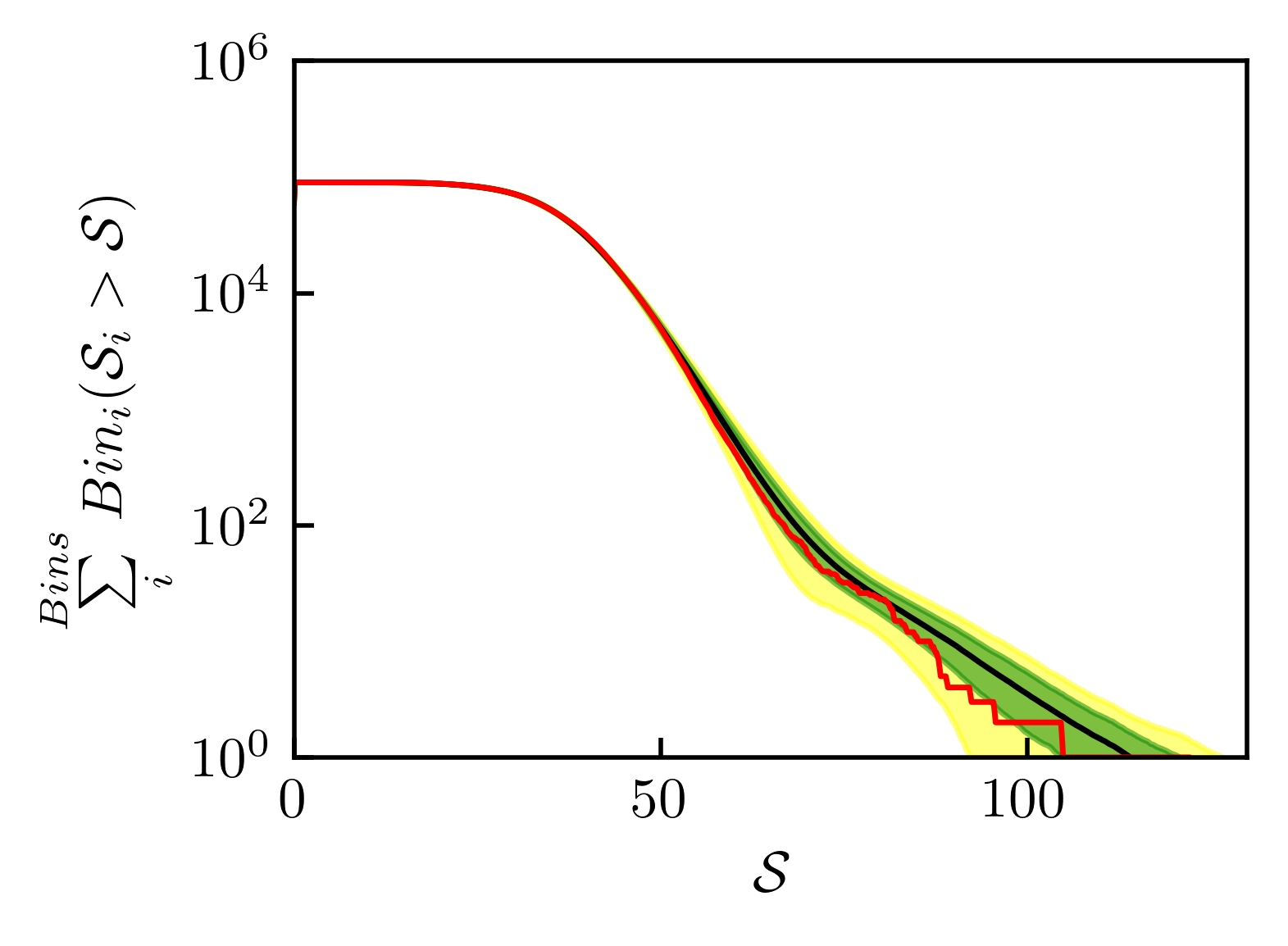}
        \caption{The measured cumulative distribution (red) compared to the mean from the scrambled ones (black) for the southern sky sample. In green and yellow, we respectively show the $1\sigma$ and $2\sigma$ confidence intervals of the scrambled sets. Note that the data agrees well with the null hypothesis. This means there are no significant deviations from a diffuse background assumption.}\label{fig:cumulative_down}
        \end{center}
    \end{figure}

    We follow the same procedure when analyzing a source set at the galactic center, as the previous discussions on the northern sky sample. We analyze the 10$^\circ$ band around the galactic center position, where we expect IceCube's sensitivity to be far smaller than the northern one, due to the large atmospheric background. Following \ref{app:distro}, we construct a cumulative distribution for the data (red) compared to the one (green) and two (yellow) sigma bands of the background model (black line) in Figure \ref{fig:cumulative_down}. This results in a p-value of approximately 0.8 for the data being background like.
    We then inject additional sources in this band, randomly placed, to construct the limits on multiple sources similar to the galactic center.

\end{document}